\documentclass[preprint2]{aastex}
\usepackage{emulateapj5}
\usepackage{onecolfloat5}

\newcommand{\beq}   {\begin{equation}}
\newcommand{\eeq}   {\end{equation}}

\newcommand{\kms}   {km~s$^{-1}$}
\newcommand{\water}   {H$_2$O~}

\shortauthors{Vlemmings, Bignall \& Diamond}
\shorttitle{The water masers of NGC~3079}
\slugcomment{For the Astrophysical Journal}

\begin{document}
\twocolumn[
\title{Green Bank Telescope observations of the water masers of NGC~3079: accretion disk magnetic field and maser scintillation}
\author{ W.~H.~T.~Vlemmings\altaffilmark{1}, H.~E.~Bignall\altaffilmark{2} \& P.~J.~Diamond\altaffilmark{1}}

\begin{abstract}
 We present observations of the 22~GHz \water megamasers in the
 circumnuclear disk of NGC~3079 obtained with the Green Bank
 Telescope. The data are analyzed for circular polarization due to the
 Zeeman-induced splitting of the \water maser lines. No circular
 polarization is detected and we derive a $1\sigma$ upper limit of
 $11$~mG for the toroidal magnetic field at $\sim 0.64$~pc from the
 central black hole. This is the tightest upper limit for the magnetic
 field around a black hole to date. We use the magnetic field limit to
 derive an estimate of the mass accretion onto the central black
 hole. In addition to the polarimetric results, we present an
 observation of rapid variability in the maser lines, which we explain
 as weak interstellar scintillation. From the scintillation
 parameters, we estimate an intrinsic size of the mostly saturated
 maser features of $\sim 12$ microarcseconds. This is consistent
 with models assuming a thick, clumpy accretion disk.
\end{abstract}
]

\altaffiltext{1}{Jodrell Bank Observatory, University of Manchester, Macclesfield, Cheshire SK11 9DL, U.K.; wouter@jb.man.ac.uk}
\altaffiltext{2}{Joint Institute for VLBI in Europe, Postbus 2, 7990~AA Dwingeloo, Netherlands}

\section{Introduction}
Since AGN are thought to be powered by accretion onto a central
super-massive black hole in which magnetic fields likely play an
important role \citep[e.g.][]{Blandford82}, substantial magnetic
fields are expected to be present throughout the entire accretion disk.
Observations of \water maser circular polarization induced by
Zeeman-splitting are an extremely useful method of probing the
magnetic field in regions where H$_2$O masers occur. Weak Zeeman
splitting was initially observed in interstellar \water masers using
the Effelsberg 100~m telescope by \citet{Fiebig89} and has since been
used to determine the strength and structure of magnetic fields in
star-forming regions \citep[e.g.][]{Sarma02,Vlemmings06b} and in the
envelopes of late-type stars \citep[e.g.][]{Vlemmings02}. 

Recently, \citet{Herrnstein98} and \citet{Modjaz05} (hereafter M05)
have used both the Very Large Array (VLA) and Green Bank Telescope (GBT)
in an attempt to measure the circular polarization of
the \water megamasers in the circumnuclear disk of NGC~4258. The
recent observations of M05 have placed stringent upper limits on
the circumnuclear magnetic field of NGC¬4258. Here we present GBT
observations of the \water masers in the circumnuclear disk of
NGC~3079 that were aimed at measuring the \water maser Zeeman splitting
due to the magnetic field in the disk.

NGC~3079 is an almost edge-on spiral galaxy at approximately 16~Mpc \citep{Sofue99}.
It shows either Seyfert-2 or LINER activity and there is strong
evidence from X-ray data that it contains an active galactic
nucleus (AGN) \citep[e.g.][~and references
  therein]{Irwin03,Kondratko05}.  NGC~3079 also contains one of the
most luminous 22~GHz \water megamasers known to date
\citep[e.g.][]{Henkel84}. Like that in NGC~4258, the \water maser
emission of NGC~3079 has been interpreted as originating in a
circumnuclear disk \citep[e.g.][]{Trotter98, Sawada00, Yamauchi04, Kondratko05}
, although the exact disk models proposed differ between authors.
\citet{Kondratko05} (hereafter K05) recently produced the first map of
the full extent of the 22~GHz \water maser emission of NGC~3079 (between $V_{\rm LSR}\approx 880~$\kms and $1400~$\kms) and
propose a model in which the accretion disk of NGC~3079 is thick, clumpy,
flaring and undergoing star-formation. NGC~3079 is also the site of the
first detected extragalactic submillimeter \water masers, which also
originate from the nuclear region \citep{Humphreys05}.

In addition to the measurement of Zeeman splitting, the GBT
observations have allowed us to study the short-term variability of
the masers. Previously, line flux variability of several tens of per
cent on timescales of minutes has been observed for 22~GHz \water
megamasers of the Circinus galaxy \citep{Greenhill97, McCallum05}. As
the timescale of intrinsic maser variations are at least an order of
magnitude larger, the Circinus maser variability is likely due to a
combination of weak and diffractive interstellar scintillation
\citep{McCallum05}. Here we present the first measurements of short
timescale variability of the NGC~3079 \water megamasers and discuss
the variability in the context of interstellar scintillation.

\section{Observations \& Data Reduction}

The observations of the \water megamasers in NGC~3079 were carried out
at the NRAO\footnote{The National Radio Astronomy Observatory (NRAO)
  is a facility of the National Science Foundation operated under
  cooperative agreement by Associated Universities, Inc.} GBT between
April 10 and April 19 2006. At 22.2~GHz the full-width at half maximum
(FWHM) beamwidth of the GBT is $\sim 33''$, several orders of
  magnitude larger than the size of the \water maser region,
  which is contained within approximately 30~mas. The GBT spectrometer
was used with a bandwidth of 200~MHz and 16,384 spectral
channels. This resulted in a channel spacing of $0.164$~\kms~ and a
total velocity coverage of $2700$~\kms~, which was centered on $V_{\rm
  LSR}=1116$~\kms. Furthermore, the data were taken with the
dual-polarization receiver of the lower $K$-band (frequency range:
18.0-22.5~GHz) using the total power nod observing mode. The two
beams have a fixed separation of $178.8''$ in the azimuth direction
and a cycle time of 2 minutes was sufficient to correct for
atmospheric variations. As a result, one beam of the telescope was
always pointing at the source while the other beam was used for
baseline correction.  The total observing time was 30.2 hours of which
20.9 hours were spent on source. Once every $\sim1.2$~hr point and
focus observations were done on J0958+655. This source was also used
as flux calibrator. The pointing corrections resulted in a pointing
accuracy typically better than $\sim2''$. System temperatures ranged
between $35$ and $55$~K, although an observing block of $\sim 5$~hr
suffered from bad weather and as a result had a higher system
temperature of $\sim 65$~K.

\begin{figure}[t!]
\epsscale{1.0} 
\plotone{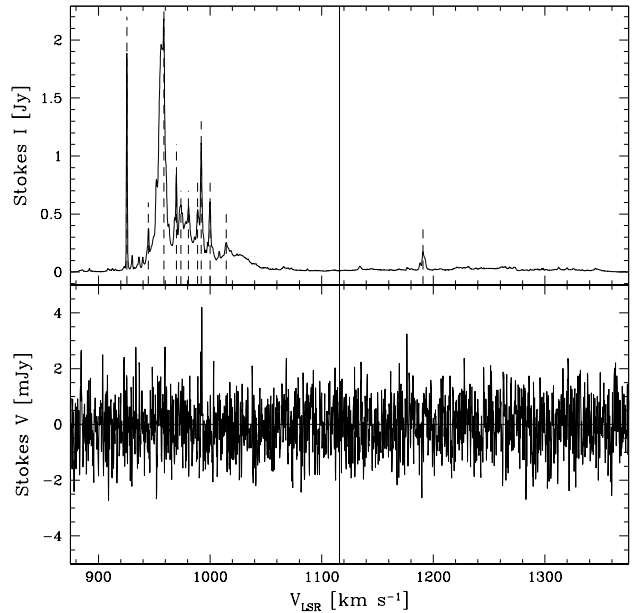}
\caption[spec]{Average total intensity (I, top) and circular polarization (V, bottom) spectrum of the NGC~3079 \water megamasers. The solid vertical line at $V_{\rm LSR}=1116$~\kms~ indicates the systemic velocity. The dashed vertical lines indicate the maser features for which the variability parameters are given in Table~\ref{Table:res}.}
\label{Fig:spec}
\end{figure}

\begin{figure}[t!]
\epsscale{1.0} 
\plotone{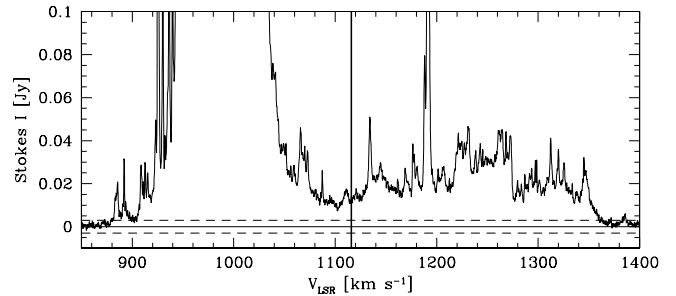}
\caption[faintspec]{Rescaled hanning smoothed total intensity spectrum of the NGC~3079 \water megamasers highlighting the faintest detected maser flux. The solid vertical line indicates the systemic velocity. The dashed horizontal lines give the $3\sigma$ rms level of the hanning smoothed spectrum.}
\label{Fig:specfaint}
\end{figure}

For data reduction the GBT IDL\footnote{http://gbtidl.sourceforge.net}
software package was used. The observations of J0958+655 yielded a
gain-curve over the entire observation session which was linearly
interpolated between calibrator observations and applied to the
NGC~3079 data. Over the full observation run, the gain does not change
by more than $10\%$, while the variation over several hours in
  the flux density of J0958+655 is $<5\%$. Assuming that J0958+655 is
not circularly polarized and that the circular polarization of the
continuum emission of NGC~3079 is negligible, both the calibrator and
the source data were used to determine a constant gain offset between
the RCP and LCP data. It was found that the RCP gain needed to be
multiplied by a factor $0.945$ to produce a Stokes V $(=({\rm
  RCP}-{\rm LCP})/2)$ spectrum without signal.  To remove residual
baseline variations due to atmospheric and instrumental effects as
well as removing the continuum emission from NGC~3079, we subtracted a
baseline determined using the spectral channels free of maser
emission.  We estimate the absolute flux calibration to be accurate to
$\sim 10\%$. As the magnetic field determination depends on the S/N
and not the absolute flux levels, this uncertainty does not influence
the field strength determination. Finally, we produce source spectra
in Stokes I and V, seen in Fig.~\ref{Fig:spec}, as well as RCP and LCP
for the total combined data set. We find, from the spectral channels
without signal, that we reach a $1\sigma$ noise level of $1.4$~mJy for
I and V, while for RCP and LCP the noise level is $2.1$~mJy. As
  the GBT observations presented here constitute hitherto the most
  sensitive observations of the \water megamasers of NGC~3079,
  Fig.~\ref{Fig:specfaint} shows a rescaled, hanning smoothed, total
  power spectrum highlighting the faint maser emission. The $1\sigma$
  noise level in this hanning smoothed spectrum is $0.9$~mJy.

\section{Results \& Discussion}

\subsection{Maser Saturation}

Theoretical analysis in \citet[e.g][]{Nedoluha92} has shown that the
degree of circular polarization of the 22~GHz \water masers depends on
the level of maser saturation. The 22~GHz \water masers begin to
saturate when $R/\Gamma \gtrsim 1$~s$^{-1}$ and approach full
saturation when $R/\Gamma \gtrsim 100$~s$^{-1}$, where $R$ is the rate
of stimulated emission and $\Gamma$ is the maser decay rate.
Following the analysis in M05, we find for the stimulated emission
rate of the \water masers of NGC~3079 at $16$~Mpc \beq R = 250
({{S_\nu} \over {1~{\rm Jy}}}) ({{L} \over {10^{16}~{\rm
      cm}}})^{-2}~{\rm s}^{-1}.  \eeq Here we assume a cylindrical
maser model with $L$ the maser length and $S_\nu$ the maser flux
density. The clumpy masing disk model from K05 predicts clump sizes
between $0.001$ and $0.006$~pc. Taking these sizes as limits on the
maser length $L$, we find $70(S_\nu / 1~{\rm Jy})<R<2600(S_\nu /
1~{\rm Jy})$~s$^{-1}$. The lifetime of the IR transitions
involved in the masing process $\Gamma=1$~s$^{-1}$, and the typical
maser features are between $0.25$ and $0.5$~Jy; we thus conclude that most
of the \water masers of NGC~3079 are at least partially saturated and
that several of the masers are even fully saturated. This is in
contrast to the mostly unsaturated masers of NGC~4258 (M05) and to the
prediction of $R/\Gamma \lesssim 1$ from numerical simulations by
\citet{Watson00}. It is worth noting that the models from
\citet{Watson00} are specifically tailored for the thin accretion disk
of NGC~4258 and are thus unlikely to be applicable to the thick,
clumpy accretion disk of NGC~3079. For high saturation, the spectral
FWHM linewidth is expected to be $\Delta v\approx 1.3$~\kms~
\citep[e.g][]{Nedoluha92, Vlemmings02} for masing gas at intrinsic
temperatures $T\sim300$~K. In Table~\ref{Table:res}, we notice that
although the majority of the maser lines reach the expected width,
several of the strongest lines have a $\Delta v\sim 0.9$~\kms. This
either indicates that these lines are only partly saturated, that the
intrinsic thermal line width of these lines is smaller, or a
combination of these effects, thus highlighting various different conditions
throughout the maser disk.

\begin{deluxetable}{lcccccc}
\tablecolumns{7}
\tablewidth{0pc} 
\tablecaption{Magnetic field upper limits\label{Table:mag}}
\tablehead{ 
\colhead{Maser line(s)} & \colhead{Fitted velocity interval} & \colhead{$\Delta_v$} &
\colhead{$I_0$} & \colhead{$B_{\rm ||}$} &  \colhead{$\sigma_{B_{\rm ||}}$} & \colhead{$|B_{\rm ||}^{\rm lim}|$}  \\
\colhead{} & \colhead{(\kms)} & \colhead{(\kms)} &
\colhead{(Jy)} & \colhead{(mG)} & \colhead{(mG)} & \colhead{(mG)} 
}
\startdata
$925.503$~\kms~ feature & $923.5$--$927.5$ & 1.913 & 0.81 & - & 16 & 48  \\
\hline
Red-shifted lines & $1175$--$1375$ & - & - & 349 & 269 & 807 \\
Blue-shifted lines & $900$--$1100$ & - & - & -9 & 11 & 33 \\
\enddata 
\end{deluxetable} 

\subsection{The Magnetic Field}

\begin{figure}[t!]
\epsscale{1.0} 
\plotone{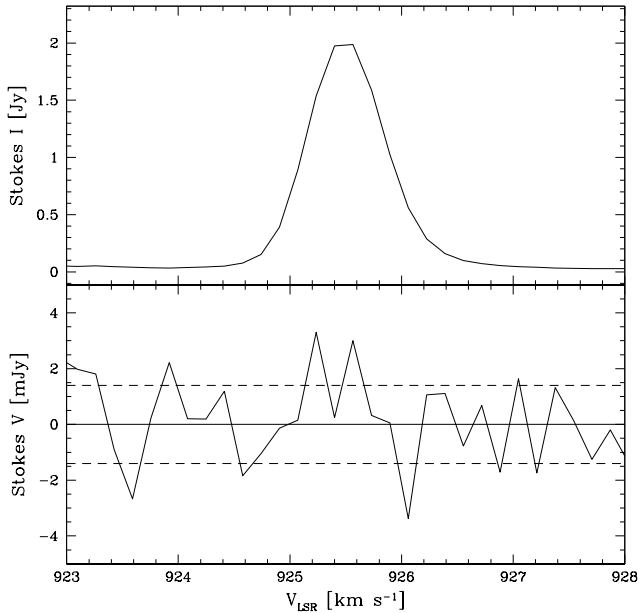}
\caption[sspec]{Average total intensity (I, top) and circular polarization (V, bottom) spectrum of the $925.5$~\kms~ blue-shifted \water maser feature of NGC~3079. The dashed horizontal lines in the V-spectrum indicate the $\pm1\sigma$ errors.}
\label{Fig:sspec}
\end{figure}

As the \water maser lines in the spectrum of NGC~3079 are very
blended, we used the RCP-LCP cross-correlation method described in M05
to determine the Zeeman splitting and corresponding magnetic field
strength for both the red- and blue-shifted masers. This
cross-correlation method is specifically suited for blended maser
lines of non-paramagnetic maser species, assuming the magnetic field
is mostly constant across the spectrum, and it has been used to
determine the upper limit of the magnetic field strength from the
22~GHz \water maser polarization spectra of NGC~4258. As described in
M05, the cross-correlation method yields the velocity shift $q$ due to
the Zeeman effect, which is directly related to the magnetic field
strength along the line of sight $B_{\rm ||}$ by \beq B_{\rm ||} =
{{q} \over {\sqrt{2}A_{F-F'}}}.  \eeq Here $A_{F-F'}$ is the
coefficient that describes the relation between the circular
polarization and the magnetic field for a transition between a high
($F$) and low ($F'$) rotational energy level. The coefficient depends
both on the intrinsic maser thermal width and the maser saturation
level \citep{Nedoluha92, Vlemmings02}. Here we have used
$A_{F-F'}=0.020$~\kms~G$^{-1}$, as derived by \citet{Nedoluha92} for
the merging of the three dominant 22~GHz \water maser hyperfine
transitions. The results of this analysis are given in
Table~\ref{Table:mag}, where the rms error in velocity shift $q$,
determined as described in the appendix of M05, yields the error in
$B_{\rm ||}$. As the red-shifted masers are weak, the $1\sigma$ error
on the $B_{\rm ||}$-field determination is large, and as no
significant magnetic field detection is made, we determine the
$3\sigma$ upper limit to be $807$~mG. The error on the magnetic field
in the blue-shifted masers is significantly smaller, but still no
significant Zeeman splitting was detected. Here the $3\sigma$ upper
limit is $33$~mG, hitherto the tightest upper limit on the magnetic
field in an accretion disk.

In addition to the cross-correlation method for the heavily blended
maser spectrum, an isolated \water maser line at $V_{\rm
LSR}=925.5$~\kms~ also allows us to derive the magnetic field using
the regular Zeeman method \citep{Vlemmings02}. Here the magnetic field
is determined using \beq P_V = {{(V_{\rm max}-V_{\rm min})} \over
{I_{\rm max}}} = 2 A_{F-F'} B_{\rm ||} / \Delta v, \eeq where the
fractional circular polarization $P_V$ is determined from the maximum
and minimum circular polarization $V_{\rm max}$ and $V_{\rm min}$ and
the peak total intensity $I_{\rm max}$. The coefficient $A_{F-F'}$ is
described above and is again taken to be $0.020$~\kms~G$^{-1}$, and
$\Delta v$ is the maser FWHM. As seen in Fig.~\ref{Fig:sspec}, no
significant detection is made in the circular polarization
spectrum. Using the calculation of the rms error limit as described in
\citet{Vlemmings06b}, we find a $3\sigma$ upper limit of
$B_{\rm ||}=48$~mG, similar to the upper limit determined on the
blue-shifted masers with the cross-correlation method.

Assuming a standard accretion disk model, the magnetic field
measurements at the blue and red-shifted maser features probe the
toroidal magnetic field $B_\phi$ at an average distance to the central
black hole of $\sim 0.64$~pc.  Since we found that the masers appear
to be be either partially or fully saturated, the measured magnetic
field strength could overestimate the true field strength by up to a
factor of $4$ \citep[][~fig. 7]{Vlemmings02}. However, velocity
gradients along the maser path can cause the true magnetic field
strength to be underestimated by up to a factor of two
\citep{Vlemmings06a}. Additionally, the blending of maser lines in
single-dish measurements can also lead to further underestimation of
the actual field strength by a factor of two \citep{Sarma01}. As we
determined a similar upper limit on the isolated maser line at $V_{\rm
  LSR}=925.5$~\kms~ as on the blended blue-shifted masers, we adopt
the conservative $3\sigma$ upper limit of $B_\phi = 33$~mG for the
actual toroidal magnetic field strength. Using the model predictions
for the magnetic field configuration in an accretion disk from
\citet{Hawley96} as described in M05, $B_\phi$ dominates the total
magnetic field and $B_{\rm tot}\approx(16/10)^{1/2} B_\phi$. We thus
take the total magnetic field strength to be $B_{\rm tot}\lesssim
40$~mG at a radius of $0.64$~pc. Such a field strength value falls in the
lower range of the magnetic field strengths observed for the \water
masers in Galactic starforming regions \citep[e.g.][]{Fiebig89,
  Sarma02, Vlemmings06b}. Comparing to the limits found for NGC~4258
while taking into account the difference in the limit calculations,
the magnetic field limit of NGC~3079 found in this paper is $\sim 2.5$
times lower at $0.64$~pc than that at $0.14$~pc in NGC~4258.

\subsubsection{Mass Accretion Rate}

The mass accretion rate in the disk of NGC~3079 can be estimated from
the magnetic field strength using the relation between the thermal
pressure and the magnetic pressure within the accretion disk. Adapting
formula 7 from M05 to the case of NGC~3079 gives for the mass accretion rate $\dot{M}$
\begin{eqnarray}
\dot{M} \lesssim & 10^{-1.5}~{{0.6\beta}\over{1+\beta}}~({{B_{\rm tot}}\over{40~{\rm mG}}})^2~({{T_k}\over{400~{\rm K}}})^{1/2} \nonumber\\
& \times({{r}\over{0.64~{\rm pc}}})^3~({{M_{\rm BH}}\over{2\times10^6 M_{\odot}}})^{-1}~M_{\odot}{\rm yr}^{-1}
\end{eqnarray}
Here $T_k$ is the temperature of the maser gas, $r$ the distance to
the central black hole and $M_{\rm BH}$ the black hole mass. The ratio
between gas and magnetic pressure $\beta$ is estimated from local
three-dimensional MHD calculations to be $\langle\beta\rangle\approx
60$ \citep{Hawley96}. 

Thus for our total magnetic field upper limit of $B_{\rm tot}=40$~mG
we find $\dot{M}\lesssim 10^{-1.7} M_{\odot}$yr$^{-1}$ at $r=0.64$~pc, assuming a smooth magnetic field configuration. As this estimate
is based on the average magnetic field strength across the maser
region, a much more complex magnetic field structure would lead to a higher upper limit by at most an order of magnitude. We can
compare this upper limit to the estimate of the mass accretion rate
found in K05 for their model of a thick, clumpy
accretion disk. They find an accretion rate of $0.007 L_{\rm bol,10}~
M_{\odot}$yr$^{-1}$, where $L_{\rm bol,10}$ is the AGN bolometric
luminosity in units of $10^{10} L_{\odot}$. With an estimate of the
AGN bolometric luminosity from X-ray data of $\sim 5\times 10^9$--$5\times10^{10}
L_{\odot}$ this indicates $10^{-2.5} < \dot{M} < 10^{-1.5}
M_{\odot}$yr$^{-1}$. Thus our result for the mass accretion rate is
fully consistent with the K05 model and implies an AGN bolometric
luminosity of $\lesssim 2.9\times10^{10} L_{\odot}$, which in turn implies an
Eddington ratio of $\lesssim 0.4$.

\subsection{Variability}

\begin{deluxetable}{lcccccc}
\tablecolumns{7}
\tablewidth{0pc} 
\tablecaption{Variability Parameters of the NGC~3079 \water Masers Lines\label{Table:res}}
\tablehead{ 
\colhead{Velocity} & \colhead{$\Delta v$} & \colhead{$\langle S \rangle$} &
\colhead{$\sigma_{\rm S}$} & \colhead{$\mu$} & \colhead{$T_{\rm char}$} & \colhead{Depth of} \\
\colhead{(\kms)} & \colhead{(\kms)} & \colhead{(Jy)} &
\colhead{(Jy)} & \colhead{} & \colhead{(hr)} & \colhead{First Minimum} 
}
\startdata
$925.503\pm0.003$\tablenotemark{a} & $0.81\pm0.01$ & 1.913 & 0.255 & 0.133 & $0.26\pm0.03$ & $-0.41\pm0.08$ \\
$944.92\pm0.02$ & $0.77\pm0.09$ & 0.378 & 0.042 & 0.110 & $0.34\pm0.04$ & $-0.30\pm0.09$ \\
$958.660\pm0.006$ & $0.89\pm0.02$ & 2.195 & 0.197 & 0.090 & $0.27\pm0.04$ & $-0.03\pm0.06$ \\
$969.84\pm0.01$ & $1.05\pm0.03$ & 0.847 & 0.089 & 0.105 & $0.27\pm0.05$ & $-0.32\pm0.10$ \\
$973.53\pm0.04$\tablenotemark{b} & $3.4\pm0.2$ & 0.584 & 0.050 & 0.086 & $0.40\pm0.10$ & $-0.13\pm0.09$ \\
$980.89\pm0.03$ & $1.45\pm0.14$ & 0.576 & 0.047 & 0.081 & $0.44\pm0.07$ & $-0.21\pm0.09$ \\
$988.93\pm0.04$\tablenotemark{b} & $1.4\pm0.2$ & 0.539 & 0.045 & 0.090 & $0.31\pm0.05$ & $-0.13\pm0.09$ \\
$992.16\pm0.01$  & $1.30\pm0.06$ & 1.122 & 0.097 & 0.087 & $0.34\pm0.03$ & $0.09\pm0.07$  \\
$999.89\pm0.01$  & $1.34\pm0.06$ & 0.612 & 0.083 & 0.136 & $0.35\pm0.03$ & $-0.28\pm0.08$ \\
$1014.45\pm0.04$\tablenotemark{b} & $3.3\pm0.2$ & 0.256 & 0.025 & 0.096 & $0.38\pm0.10$ & $-0.08\pm0.15$ \\
$1190.65\pm0.03$ & $1.48\pm0.05$ & 0.177 & 0.025 & 0.139 & $0.37\pm0.05$ & $-0.27\pm0.18$ \\
\enddata 
\tablenotetext{a}{For the variability analysis the linear flux decrease has been removed.}
\tablenotetext{b}{Heavily blended feature.}
\end{deluxetable} 

\begin{figure}[t!]
\epsscale{1.0} 
\plotone{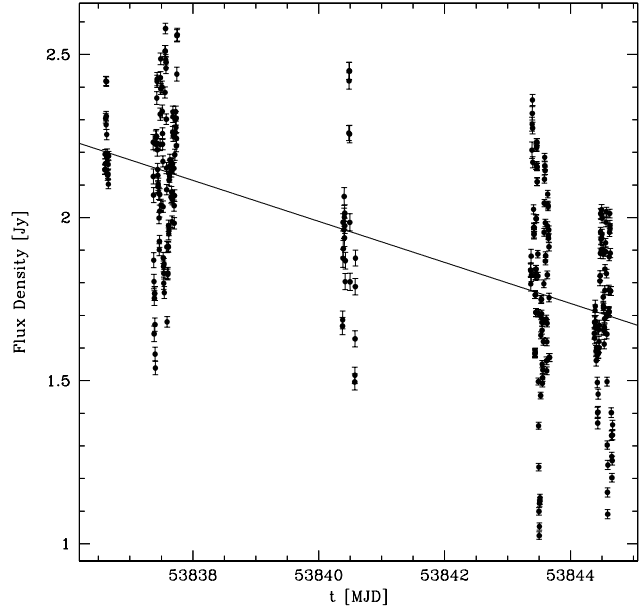}
\caption[dec]{The light curve of the $925.503$~\kms~ blue-shifted \water maser feature over the entire observation run. The thick solid line is a weighted least square fit to a simple linear decrease in flux density and corresponds to a decrease of 64~mJy per day.}
\label{Fig:dec}
\end{figure}

\begin{figure*}[t!]
\epsscale{2.0} 
\plotone{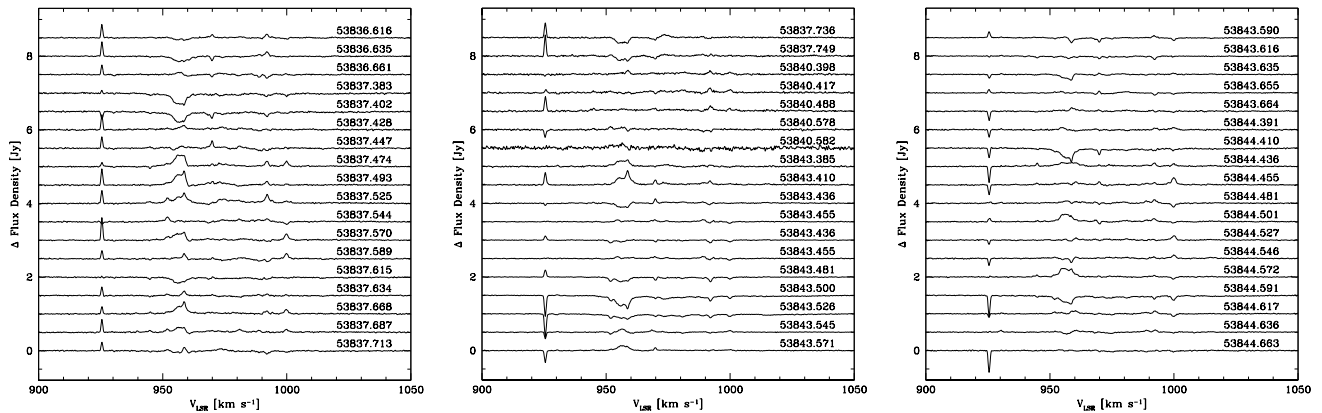}
\caption[var]{The variability of the blue-shifted \water masers of NGC~3079. The spectra are averaged over 24 minutes, corresponding to 6 observing scans. The average total intensity I spectrum from Fig.~\ref{Fig:spec} has been subtracted for each of the spectra which are labeled with the MJD of the observation. Each spectrum is off-set by $0.5$~Jy and the rms noise in the spectra is $\sim 3.4$~mJy.}
\label{Fig:varcomb}
\end{figure*}

Although the peak flux has varied by up to $~50\%$, the overall shape
of the total intensity spectrum in Fig.~\ref{Fig:spec} is strikingly
similar to the earlier spectrum from 2003 seen in figure 3 of
K05. The only major difference is the appearance of a
strong, somewhat isolated blue-shifted feature at $V_{\rm
  lsr}=925.5$~\kms. As shown in Fig.~\ref{Fig:dec}, this feature, in
contrast with the other maser features, shows a steady decline in
flux. Our time baseline is not sufficient to determine a non-linear
flux decrease. Using an error weighted, least squares fit to our
individual observing scans of 4 minutes yields a decrease in flux of
$64$~mJy/day. There are several possible mechanisms that can explain
such gradual, time-variable behavior \citep{Maloney02}. One
possibility is a time-variable maser pump. The \water megamasers could
be powered by X-ray emission from the AGN, as discussed by
\citet{Neufeld94}. Variations in the AGN would then result in
variations in the \water maser flux. However, although the location of
the $V_{\rm LSR}=925.5$~\kms~ with respect to the other maser features
is unknown, one would expect correlated flux variation in at least
some of the other masers lines. Another more likely explanation is,
that the variability is caused by the amplification of background
emission from another maser feature. As the disk of NGC~3079 is
thought to be strongly inhomogeneous, the maser emission is narrowly
beamed along the line-of-sight by the overlap of masing regions with
similar Doppler velocities. As discussed in K05, chance alignments of
masing regions will then result in variability at timescales of severals days or longer.

\subsection{Scintillation}

\begin{figure}[t!]
\epsscale{1.0} 
\plotone{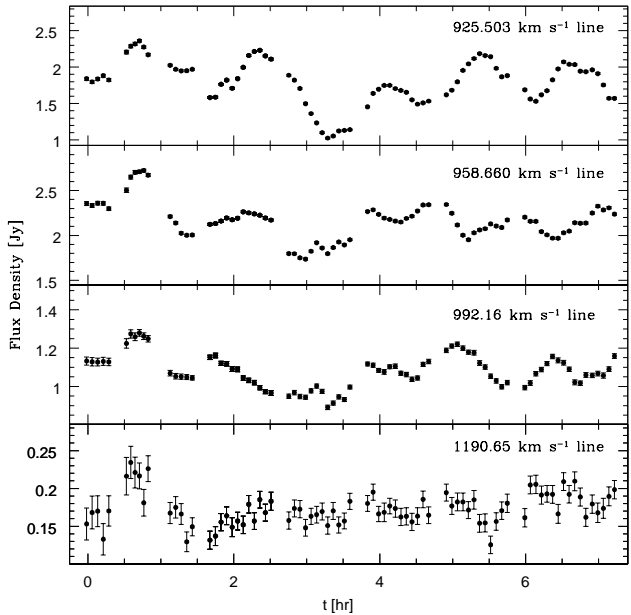}
\caption[var]{Light curves of the three strongest blue-shifted \water
  masers lines of NGC~3079 as well as that of the strongest
  red-shifted line (bottom) observed on April 17th and 18th 2006. Data
  points are plotted for every 4 minute observing scan. In the top two
  panels, the error bars are smaller than the symbol size. The typical
  $1\sigma$ error bar is $\sim 18$~mJy. The light curve of the maser feature at $925.503$~\kms has been corrected for the linear flux decrease shown in Fig.~\ref{Fig:dec}.}
\label{Fig:var}
\end{figure}

In addition to the longterm variability seen in the feature at $V_{\rm
  LSR}=925.5$~\kms, the maser features of NGC~3079 also vary on
timescales of several tens of minutes as seen in
Fig.~\ref{Fig:varcomb}.  Such rapid variability has been previously
observed for the \water megamasers of Circinus \citep{Greenhill97,
  McCallum05}. As seen in Fig.~\ref{Fig:var}, in which 7 hours of
  observation on the strongest red- and blue-shifted masers are shown,
  the observed variations reach the level of $50\%$ within the
  hour. This is similar to the flux variation seen in Circinus.
  Although small variations in time due to incorrect antenna gain
  corrections might still be present, the hour to hour gain variations
  determined from the observations of J0958+55 are not expected to be
  $>5\%$. However, as the gain corrections using J0958+55 are only
  performed every 1.2 hours after correcting the telescope pointing,
  accumulated pointing uncertainties and elevation effects can cause a
  global gain variation of up to $\approx 10$--$25\%$ between two
  consecutive observing blocks. While such large variations only occur
  6 times during our total observation run, the most extreme of these
  instances can be seen during the first hour of observations
  presented in Fig.~\ref{Fig:var}. These occasions are found to have
  no significant effect on our determination of the variability
  characteristics. An additional indication that the observed
  variability is not due to calibration errors is the fact that we
  find no significant correlation in the variability of the different
  maser lines.

\begin{figure}[t!]
\epsscale{1.0} 
\plotone{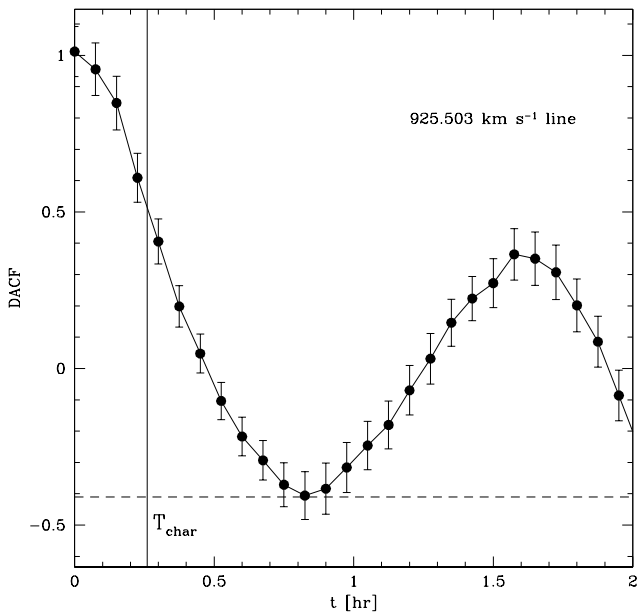}
\caption[acf]{Discrete, normalized autocorrelation function (DACF) of the $925.503$~\kms~ maser line. The lags are binned in 4.5~min intervals. The DACF and the $1\sigma$ error bars are calculated following \citet{Edelson88}. The solid vertical line indicates the characteristic variability timescale $T_{\rm char}$, where DACF$=0.5$ and the dashed horizontal line is the first negative.}
\label{Fig:acf}
\end{figure}

To characterize the variability we have determined the discrete
autocorrelation function (DACF) following \citet{Edelson88}. The
DACF of one of the maser lines is shown in
Fig.~\ref{Fig:acf}. Following \citet{Rickett02}, we define the
characteristic timescale for variability, $T_{\rm char}$, as the
time-lag where the DACF falls to $0.5$. We further determine the depth
of the first DACF minimum and the modulation index $\mu=\sigma/\langle
S \rangle$. Here $\langle S \rangle$ is the mean peak flux of the
maser line and $\sigma$ the rms of the variability. The variability
parameters are given in Table~\ref{Table:res} along with the local standard of rest (LSR)
velocity, $V_{\rm LSR}$, and the full-width half-maximum ($\Delta
v$). The velocity and $\Delta v$ are determined from fitting Gaussian
profiles to the maser spectrum. However, many of the maser features
are heavily blended, increasing the uncertainties in the velocity and
$\Delta v$ determinations.  We find that $T_{\rm char}$ ranges from
$\sim 900$ to $\sim 1600$~s.

As shown in \citet{McCallum05}, the variability of $\sim 700$~s of the
Circinus \water masers is likely due to a combination of weak and
diffractive scintillation. However, NGC~3079 is located at much higher
galactic latitude ($b=+48.36^\circ$) than Circinus, for which the line
of sight passes through the galactic plane. From the Galactic electron
density model NE2001 \citep{Cordes02}, we find at the position of
NGC~3079 that the predicted transition frequency between weak and
strong scintillation is $\nu_0=8.8$~GHz. The model uncertainties of
NE2001 are significant at high latitude and for individual
lines-of-sight, however NGC~3079 can reasonably be expected to be in
the weak scintillation regime at 22.2~GHz.

Using formula 6 from \citet{Walker98} for the frequency dependence of
modulation index for a point source in the weak scintillation regime,
we find from our observations, with a typical modulation index
$\mu=0.1$, the implied transition frequency between weak and strong
scintillation is $\nu_0=4.4$~GHz. The transition frequency would be
higher if the maser size invalidates the point source
approximation. If the angular source size $\theta_S$ is larger than
the angular size of the first Fresnel zone $\theta_F$ at the distance
of the scattering screen, then the modulation index is decreased. At
22.2 GHz, $\theta_F=59(z/1 {\rm pc})^{-1/2}~\mu$as, where $z$ is the
distance of the scattering screen. Assuming saturation, cylindrical
masers are beamed by a factor of $\sim3$ \citep{Elitzur94}. As the
clump sizes in the maser disk are estimated in K05 to be between
$0.001$ and $0.006$~pc, this yields for the beamed masers, at
$16$~Mpc, an angular size between $4$ and $25~\mu$as. As several maser
features are only partly saturated, the beaming is more pronounced,
and the angular sizes are even smaller. Thus, taking a typical maser
size of $\theta_S=10~\mu$as, the scattering screen would need to be at
$z\sim 150$~pc to satisfy $\theta_S/\theta_F\approx 2.3$ if the
transition frequency is in fact close to the NE2001 model 
$\nu_0=8.8$~GHz. It is interesting to note that our calibration
source J0958+655 (B0954+658), at an angular separation of $9.9^{\circ}$
from NGC~3079, showed the prototypical extreme scattering event in
1980-81 \citep{Fiedler87}, which was suggested by \citet{Fiedler94} as
being possibly associated with the edge of Galactic Loop III at a
distance of 145 pc.

However, a distant screen implies a long scintillation timescale. 
At 22.2~GHz, the variability timescale for a point source is $t_F=8100~
V_{\rm ISM}^{-1} z^{1/2}$, with $V_{\rm ISM}$ being the transverse speed of the scattering
screen with respect to the source in \kms. For $\theta_S>\theta_F$,
the characteristic variability timescale $T_{\rm char}=t_F
(\theta_S/\theta_F)$. Thus for a screen at $z=150$~pc and
$\theta_S/\theta_F = 2.3$, $T_{\rm char} = 2.3~10^5~
V_{\rm ISM}^{-1}$~s$^{-1}$. To reconcile this with the typical
observed timescale of $1200$~s requires $V_{\rm ISM}\approx 190$~\kms, which
is much larger than the expected value of $V_{\rm ISM}\approx 40$~\kms
corresponding to the velocity of the Earth with respect to the LSR in
the direction of NGC~3079 at the date of the observation.

Alternatively, the observed rapid variations and lower transition
frequency compared to the NE2001 model prediction could result from
scattering in a nearby screen, as has been found for the handful of
extremely rapid ``intra-hour'' scintillating quasars
\citep{Dennett-Thorpe00,Rickett02,Bignall03}.  Although there are some
exceptions, the maser lines in Table~\ref{Table:res} have a tendency
to display the smallest $T_{\rm char}$ and largest $\mu$ for those
lines with the lowest $\Delta v$, which are thus thought to be only
partially saturated. This would be consistent with the corresponding
maser features being the smallest due to the increased maser
beaming. We thus expect $\theta_S\lesssim\theta_F$ for the partially
saturated sources and expect $\theta_S$ to be slightly larger than
$\theta_F$ for the fully saturated sources. Taking $T_{\rm char}\approx
1000$~s from partially saturated masers to be $t_F$, we find that the
velocity and the screen distance have to satisfy $z^{1/2} = 0.12~V_{\rm ISM}$. For $V_{\rm ISM}\approx 40$~\kms\ for a screen moving with the local standard
of rest, this indicates $z\approx 25$~pc. This in turn gives
$\theta_S\approx\theta_F=12~\mu$as, thus the maser feature sizes
expected from the scintillation are consistent with those estimated
from the saturation level and the clump sizes from K05.

Although there is significant estimation error in the DACF due to the
limited sampling of the stochastic scintillation pattern, those masers
with larger $\mu$ also tend to show deeper first minima in the
DACF. This is also consistent with the sources with lowest $\mu$,
thought to be fully saturated, having $\theta_S > \theta_F$.
\citet{Rickett02} showed that anisotropic scattering in a thin screen
produces a deep first minimum or ``negative overshoot'' in the ACF,
and as the source becomes more extended, this ``negative overshoot''
is suppressed. A deep first minimum in the DACF is also seen for the
intra-hour variable quasars and for the \water masers in Circinus
\citep{McCallum05}, suggesting that anisotropic scattering in discrete ``screens'' is a
widespread phenonemon.

There is some indication of nearby scattering along other
lines-of-sight close to NGC~3079 from the MASIV VLA Survey
\citep{Lovell03}. The MASIV Survey found that, while a large
fraction of all compact, flat-spectrum extragalactic sources vary with
modulation indices typically in the range $0.01$ to $0.1$ at 5~GHz in a 72 hour
period, only a tiny fraction ($<1$\%) have characteristic
timescales of a few hours or less. J0949+5819, the closest MASIV
source to NGC~3079 at an angular separation of $3.2^{\circ}$,
showed variations with $\mu=0.15$ on a timescale of less than a few
hours in 2002 January. This is likely to be due to scattering in a
nearby screen within a few tens of pc. J0949+5819 was the most extreme variable
observed in the first epoch of the MASIV Survey apart from the already
known intra-hour variable J1819+3845 \citep{Dennett-Thorpe00}.  Of the
nine MASIV sources within $10^{\circ}$ of NGC~3079, five sources
showed significant variability in at least three out of four observed
epochs (Lovell et al., in preparation), namely J0949+5819, J0946+5020,
J0929+5013, J0958+4725, and our calibration source J0958+655 which
was also a previously known intra-day variable (IDV) \citep{Wagner93}. 
While the fraction of nearby IDV sources is large but not exceptional, it is
notable that the variations in J0929+5013 were also unusually rapid
with a characteristic timescale less than a few hours. The presence of
several rapid variables in the region of sky near NCG~3079 is
suggestive of these sources being scattered by related structure 
in the very local Galactic interstellar medium.

The proposed scintillation model could be tested by measuring the time 
delays and correlation between the maser scintillation patterns arriving 
at two widely separated telescopes. Such a two-station time delay 
experiment, for example between the GBT and Effelsberg telescopes, should 
also allow a determination of the anisotropy in the scintillation pattern 
and its position angle as well as the scintillation velocity \citep[see e.g.][]{Bignall06}.

It is worth noting that short timescale variability of masers can also
be caused by gain variation along the maser amplification path. As the
characteristic timescale of such variability is $\sim L/c$, where $L$
is the maser length, for the \water maser emission of NGC~3079 this
implies a variability timescale of this type of $\sim10^5$~s. This is
at least two orders of magnitude longer than the variability timescale
detected here.

\section{Summary}

The full velocity range of the \water megamasers in the accretion
disk of NGC~3079 has been observed using the GBT in both RCP and LCP.
This has enabled us to determine a sensitive $1\sigma$ upper limit
of $\sim 11$~mG for the absolute poloidal magnetic field
$|B_\phi|$. This corresponds to a $3\sigma$ upper limit for the total
magnetic field in the accretion disk of $B_{\rm tot, lim}=40$~mG at
$\sim 0.64$~pc from the central black hole. This is the tightest
magnetic field upper limit in an AGN accretion disk to date. The
magnetic field upper limit corresponds to a mass accretion rate of
$\dot{M}\lesssim 10^{-1.7} M_{\odot}$yr$^{-1}$, consistent with
NGC~3079 accretion disk models and the X-ray luminosity.

In addition to the magnetic field measurements, the \water maser lines
of NGC~3079 were found to vary at characteristic timescales of the
order of $1000$~s. We conclude that this variability is due to weak
scintillation caused by a scattering screen at $\sim 25$~pc. The
scintillation timescales and modulation index imply maser feature
sizes of $\sim 12~\mu$as which corresponds to approximately $180$~AU
at the distance of NGC~3079. As the masers are found to be mostly
saturated, maser beaming considerations imply that the masers occur in
clumps with a size of $\sim 0.002$~pc, consistent with the model for a
thick, clumpy accretion disk.

\begin{acknowledgements}
This research was supported by a Marie Curie Intra-European fellowship
within the 6th European Community Framework Program under contract
number MEIF-CT-2005-010393.
\end{acknowledgements}


\begin{thebibliography}{23}
\expandafter\ifx\csname natexlab\endcsname\relax\def\natexlab#1{#1}\fi


\bibitem[Bignall et al.(2003)]{Bignall03} Bignall, H.~E., Jauncey,
  D.~L., Lovell, J.~E.~J., Tzioumis, A.~K., Kedziora-Chudczer, L.,
  Macquart, J.-P., Tingay, S.~J., Rayner, D.~P. \& Clay, R.~W.\  2003,
  \apj, 585, 653

\bibitem[Bignall et al.(2006)]{Bignall06} Bignall, H.~E., Macquart, J.-P., Jauncey, D.~L., Lovell, J.~E.~J., Tzioumis, A.~K., Kedziora-Chudczer, L.\  2006,
  \apj in press, arXiv:astro-ph/0608619

\bibitem[Blandford \& Payne(1982)]{Blandford82} Blandford, R.~D., 
\& Payne, D.~G.\ 1982, \mnras, 199, 883 

\bibitem[Cordes \& Lazio(2002)]{Cordes02} Cordes, J.~M., \& 
Lazio, T.~J.~W.\ 2002, ArXiv Astrophysics e-prints, arXiv:astro-ph/0207156

\bibitem[Dennett-Thorpe \& de Bruyn(2000)]{Dennett-Thorpe00} 
Dennett-Thorpe, J., \& de Bruyn, A.~G.\ 2000, \apjl, 529, L65 

\bibitem[{{Edelson} \& {Krolik} (1988)}]{Edelson88}
{Edelson}, R.~A., \& {Krolik}, J.~H. 1988, \apj, 333, 646


\bibitem[Elitzur(1994)]{Elitzur94} Elitzur, M.\ 1994, \apj, 422, 
751 

\bibitem[{{Fiebig} \& {G\"usten}(1989)}]{Fiebig89}
{Fiebig}, D. \& {G\"usten}, R. 1989, \aap, 214, 333

\bibitem[Fiedler et al.(1987)]{Fiedler87} Fiedler, R.~L., 
Dennison, B., Johnston, K.~J., \& Hewish, A.\ 1987, \nat, 326, 675 

\bibitem[Fiedler et al.(1994)]{Fiedler94} Fiedler, R., Pauls, T., 
Johnston, K.~J., \& Dennison, B.\ 1994, \apj, 430, 595 

\bibitem[Greenhill et al.(1997)]{Greenhill97} Greenhill, L.~J., 
Ellingsen, S.~P., Norris, R.~P., Gough, R.~G., Sinclair, M.~W., Moran, 
J.~M., \& Mushotzky, R.\ 1997, \apjl, 474, L103

\bibitem[Hawley et al.(1996)]{Hawley96} Hawley, J.~F., Gammie, 
C.~F., \& Balbus, S.~A.\ 1996, \apj, 464, 690

\bibitem[Henkel et al.(1984)]{Henkel84} Henkel, C., Guesten, R., 
Downes, D., Thum, C., Wilson, T.~L., \& Biermann, P.\ 1984, \aap, 141, L1

\bibitem[Herrnstein et al.(1998)]{Herrnstein98} Herrnstein, J.~R., 
Moran, J.~M., Greenhill, L.~J., Blackman, E.~G., \& Diamond, P.~J.\ 1998, 
\apj, 508, 243 

\bibitem[Humphreys et al.(2005)]{Humphreys05} Humphreys, E.~M.~L., 
Greenhill, L.~J., Reid, M.~J., Beuther, H., Moran, J.~M., Gurwell, M., 
Wilner, D.~J., \& Kondratko, P.~T.\ 2005, \apjl, 634, L133

\bibitem[Irwin \& Saikia(2003)]{Irwin03} Irwin, J.~A., \& 
Saikia, D.~J.\ 2003, \mnras, 346, 977

\bibitem[Kondratko et al.(2005)]{Kondratko05} Kondratko, P.~T., 
Greenhill, L.~J., \& Moran, J.~M.\ 2005, \apj, 618, 618 (K05)

\bibitem[Lovell et al.(2003)]{Lovell03} Lovell, J.~E.~J., 
Jauncey, D.~L., Bignall, H.~E., Kedziora-Chudczer, L., Macquart, J.-P., 
Rickett, B.~J., \& Tzioumis, A.~K.\ 2003, \aj, 126, 1699

\bibitem[Maloney(2002)]{Maloney02} Maloney, P.\ 2002, 
Publications of the Astronomical Society of Australia, 19, 88 

\bibitem[McCallum et al.(2005)]{McCallum05} McCallum, J.~N., 
Ellingsen, S.~P., Jauncey, D.~L., Lovell, J.~E.~J., \& Greenhill, L.~J.\ 
2005, \aj, 129, 1231

\bibitem[\protect\citeauthoryear{{Modjaz}, {Moran}, {Kondratko} \& {Greenhill}}{{Modjaz} et~al.}{2005}]{Modjaz05}
{Modjaz} M.,  {Moran} J.~M.,  {Kondratko} P.~T., {Greenhill} L.~J.,  2005, \apj, 626, 104 (M05)

\bibitem[{Nedoluha} \& {Watson}(1992)]{Nedoluha92} 
Nedoluha, G.~E.~\& Watson, W.~D.\ 1992, \apj, 384, 185 

\bibitem[Neufeld et al.(1994)]{Neufeld94} Neufeld, D.~A., 
Maloney, P.~R., \& Conger, S.\ 1994, \apjl, 436, L127

\bibitem[Rickett et al.(2002)]{Rickett02} Rickett, B.~J., 
Kedziora-Chudczer, L., \& Jauncey, D.~L.\ 2002, \apj, 581, 103 

\bibitem[{{Sarma} {et~al.}(2002){Sarma}, {Troland}, {Crutcher}, \& {Roberts}}]{Sarma02}
{Sarma}, A.~P., {Troland}, T.~H., {Crutcher}, R.~M., \& {Roberts}, D.~A. 2002, \apj, 580, 928

\bibitem[\protect\citeauthoryear{{Sarma}, {Troland} \& {Romney}}{{Sarma} et~al.}{2001}]{Sarma01}
{Sarma} A.~P.,  {Troland} T.~H.,    {Romney} J.~D.,  2001, \apjl, 554, L217

\bibitem[Sawada-Satoh et al.(2000)]{Sawada00} Sawada-Satoh, S., 
Inoue, M., Shibata, K.~M., Kameno, S., Migenes, V., Nakai, N., \& Diamond, 
P.~J.\ 2000, \pasj, 52, 421 

\bibitem[Sofue(1999)]{Sofue99} Sofue, Y.\ 1999, Advances in 
Space Research, 23, 949 

\bibitem[Trotter et al.(1998)]{Trotter98} Trotter, A.~S., 
Greenhill, L.~J., Moran, J.~M., Reid, M.~J., Irwin, J.~A., \& Lo, K.-Y.\ 
1998, \apj, 495, 740 

\bibitem[{{Vlemmings}(2006)}]{Vlemmings06a}
{Vlemmings}, W.~H.~T. 2006, \aap, 445, 1031

\bibitem[{{Vlemmings} {et~al.}(2002){Vlemmings}, {Diamond}, \& {van Langevelde}}]{Vlemmings02}
{Vlemmings}, W.~H.~T., {Diamond}, P.~J., \& {van Langevelde}, H.~J. 2002, \aap,  394, 589


\bibitem[\protect\citeauthoryear{{Vlemmings}, {Diamond}, {van Langevelde} \&  {Torrelles}}{{Vlemmings} et~al.}{2006}]{Vlemmings06b}
{Vlemmings} W.~H.~T.,  {Diamond} P.~J.,  {van Langevelde} H.~J., {Torrelles} J.~M.,  2006, \aap, 448, 597

\bibitem[Wagner et al.(1993)]{Wagner93} Wagner, S.~J., et al.\ 
1993, \aap, 271, 344 

\bibitem[Walker(1998)]{Walker98} Walker, M.~A.\ 1998, \mnras, 
294, 307 

\bibitem[Watson \& Wyld(2000)]{Watson00} Watson, W.~D., \& Wyld, 
H.~W.\ 2000, \apj, 530, 207 

\bibitem[Yamauchi et al.(2004)]{Yamauchi04} Yamauchi, A., Nakai, 
N., Sato, N., \& Diamond, P.\ 2004, \pasj, 56, 605 

\end{thebibliography}
\end{document}